\documentclass{JHEP3}
\usepackage{epsf}
\usepackage{amsmath}
\usepackage{amsfonts}
\usepackage{amssymb}

\newcommand{\pa}{\partial}
\newcommand{\tr}{{\rm tr}}
\newcommand{\comment}[1]{}

\newcommand{\pasl}{\pa\kern-.55em /}

\newcommand{\ksl}{k\kern-.55em /}

\DeclareFixedFont{\xiiss}{OT1}{cmss}{m}{n}{12}
\DeclareFixedFont{\ixss}{OT1}{cmss}{m}{n}{9}
\DeclareFixedFont{\cmrnine}{OT1}{cmr}{m}{n}{9}
\newcommand{\field}[1]{\mathbb{#1}}
\newcommand{\BC}{{\field C}}

\newcommand{\BZ}{{\field Z}}

\newcommand{\CCs}{\hbox{\ixss C\kern-.4emI}}
\newcommand{\ZZs}{\hbox{\ixss Z\kern-.4emZ}}

\newcommand{\CA}{{\cal A}}

\newcommand{\CZ}{{\cal Z}}

\newcommand{\ZA}{\CZ\CA}

\newcommand{\CP}{{\BC\field P}}

 \newcommand{\myfig}[3]{\begin{figure}[ht]
\begin{center}
\leavevmode
\epsfxsize=#2cm
\epsfbox{#1}
\end{center}
\caption{#3}
\label{fig:#1}
\end{figure}}

\preprint{hep-th/0201093}

\title{Reverse geometric engineering of singularities}

\author{David Berenstein \\
School of Natural Sciences,
Institue of Advanced Study,
Einstein Drive, NJ 08540\\
Email: \email{dberens@ias.edu} }

\abstract{One can geometrically engineer supersymmetric field theories
theories by
placing D-branes at or near singularities. The opposite  
process is described, where one can reconstruct the singularities
from quiver 
theories. The description is in terms of a noncommutative 
quiver algebra which is constructed
from the quiver diagram and the superpotential. The center of this 
noncommutative algebra is a commutative algebra, which is the ring of 
holomorphic functions on  a variety $V$. If certain algebraic conditions are
 met, then the reverse geometric engineering produces $V$ as the geometry 
that D-branes probe.
It is also argued that the identification of $V$ is 
invariant under Seiberg dualities.
}

\begin{document}

\section{Introduction}

String theory has solutions 
at weak coupling that correspond to propagation on geometrically 
singular spaces. In particular, one can consider Calabi-Yau 
compactifications that correspond to an ${\cal N}=(2,2)$
superconformal field theory whose target space 
geometric interpretation is of
strings propagating on a singular Calabi-Yau manifold
with a constant dilaton and without $RR$ backgrounds.

However, although the na\"\i ve geometric interpretation is 
of strings propagating on  a singular space,
the worldsheet conformal field theory is nonsingular: the singularity 
is resolved by a stringy mechanism. 

General worldsheet conformal field theories are very hard to analyze, so 
instead one
can hope that point-like D-brane probes give a good account of 
the geometry and give some new notion of geometry
where the space is smooth.
In \cite{BL4} it was proposed that the natural D-brane
 notion of smoothness of 
this space is given in terms of a {\it regular} non-commutative geometry
 and in essence 
this non-commutative geometry gives us a resolution of the commutative 
singularities.

Once we have a D-brane probe near the 
singularity, we can take the limit
of large volume Calabi-Yau and we can then take an $\alpha'\to 0$
decoupling limit, so we are left over with a supersymmetric
field theory
on the world-volume of the D-brane, which has ${\cal N} = 1$ supersymmetry.
 The precise form of 
this $\alpha'\to 0$ 
limit is the main subject of the AdS/CFT correspondence, and here it is 
interpreted as taking the low energy effective field theory of the 
D-brane on it's moduli space of vacua. By this token, the theory does
not have to be renormalizable, but we will also ignore the K\"ahler 
potential of the theory, so we will only be interested in holomorphic 
information. 

This idea provides a connection between (singular)
complex geometry and 
supersymmetric field 
theories and  it is usually called geometric engineering.
This is, given a singularity, one can construct field 
theories associated to it.
However, for most singularities it is not known how to extract the 
field theory that describes the singularity. 
A proposal has been worked out for the specific examples of
 orbifolds \cite{DM}, orbifolds with discrete torsion \cite{D,DF,BL}, 
toric singularities (without discrete torsion) 
\cite{FHH,FHH2,FHHU} and for one parameter families
of resolved  ALE singularities \cite{CIV,CKV,CIKV}, of which a special 
example
is the conifold \cite{KW}.

A few things are known about this process, at least for the type II
string theory. If we have a Calabi-Yau threefold singularity 
${\cal X}$, then placing a
collection of $N$ D3-branes
near the singularity produces a gauge field theory with 
gauge group $\prod_i U(N N_i)$, with matter transforming as the
$(N N_i, \bar {N N_j})$ for some gauge groups, and with a 
superpotential of single trace type. If we allow the $N N_i$ to be 
arbitrary
integer numbers then we say we have a configuration of fractional 
branes.

The single trace property can be understood from the 
string worldsheet point of view as having just 
 one boundary on the worldsheet.
Indeed, the
contribution of multi-boundary worldsheets is suppressed by the string
coupling which we formally take towards zero. Perturbative
 nonrenormalization
 theorems
will prevent us from generating a multi-trace superpotential 
from loop diagrams, and can only be generated 
nonperturbatively. The reconstruction of the geometry depends only on the
 classical superpotential, this is, we do not need to have 
$D3$ branes near the singularity, any lower dimensional D-brane obtained by 
dimensional reduction of the theory is just as good for reconstructing the 
geometry.

One also expects that the classical moduli space of the
 D-branes is given by
the symmetric product $Sym^N({\cal X})$ so long as there are no 
RR backgrounds which can produce 
potential terms for D-branes propagating on $X$. 
This type of potential generically 
localizes the branes on a submanifold and can give rise to Myers
type effects \cite{Myers} which makes the D-branes into extended
objects (and therefore nonlocal probes) 
 as opposed to point-like probes of the 
geometry. So, in the absence of 
RR backgrounds, we can 
recover ${\cal X}$ from the moduli
space of D-branes. It is also in this case that worldsheet 
supersymmetry
is unbroken, and that one 
can have a topologically twisted ${\cal N}=2$ topological
string theory. This topologically twisted 
theory can compute the superpotential for a collection 
of D-branes.

Similarly, one can ask the opposite question: given a field theory
that admits a large $N$ limit, with matter transforming in bifundamentals, 
a single-trace type superpotential, can we recover such 
a singular space ${\cal X}$ from this data? 
This is the question that we will ask, and we will call the process
of finding such an ${\cal X}$ reverse geometric engineering, being the 
opposite of the procedure described in \cite{KMV}. This is
 in the sense
that we usually engineer field theories by putting D-branes 
on singularities and 
ask what the low energy limit of the D-branes near the singularity 
corresponds to.

The paper is organized as follows:

In section \ref{sec:recipe} we provide an algebraic procedure for producing
 the space ${\cal X}$ from the quiver data, and the conditions under which 
${\cal X}$
 is expected to be the right answer to the reverse geometric engineering
problem.

In section \ref{sec:fam} we give a set of examples from families of
deformed $A_n$ singularities. We explicitly compute the center of the algebra
in full
detail and reproduce the moduli space of vacua with a very compact 
calculation.

Next, in section \ref{sec:nontor} we discuss a  simple
 non-toric singularity, which is not 
an orbifold either. Here we see that the techniques presented
can make sense for field theories which are interesting for
the AdS/CFT correspondence, by performing 
marginal deformations of the superpotential of known theories.

In \ref{sec:Z3} we show in a particular example
that the geometric space ${\cal X}$ is well
defined even if one performs Seiberg dualities on
the nodes. We argue that this feature is generic: 
the extracted geometry does not depend
on which dual realization of a field theory one is using,
that in some sense the variables in the center of the algebra
are gauge invariant, and therefore they do not transform under 
field theory dualities.

\section{A non-commutative algebra for each 
 quiver diagram}\label{sec:recipe}

Consider a quiver diagram with a finite number of vertices $V_i$, $i=1,
 \dots, n$
 and (directed)
arrows
$\phi^{a}_{ij}$, where the subindex labels indicate the vertices where 
the arrows begin and end. This also includes the possibilities
of arrows that begin and end on the same node.
The data of the quiver theory includes
a choice of superpotential.

Usually when we consider a field theory associated to a quiver diagram we 
specify a gauge 
group $U(N_i)$ for each vertex of the diagram, and a chiral multiplet for 
each arrow $\phi^a_{ij}$ which transforms in the $(N_i, \bar N_j)$ of the
gauge fields at both ends. If the arrow begins and ends in the same node,
 this is an adjoint superfield. In our case, the $N_i$ will be considered as 
arbitrary numbers,
which can be set to zero if we want to
and we will consider the full family of field theories associated to a 
given quiver 
diagram with arbitrary $N$.
We will assume that the quiver diagram contains
as vertices all of the topologically distinct fractional
branes that appear at a singularity. If this condition
is not met, then
we will call the quiver diagram incomplete. However,
this is a test that is
done a
posteriori if the procedure for finding the geometry where the branes 
propagate fails. One should expect in general that if one is given a 
set of conformal 
field theories labeled by $N$, or a set of theories labeled by $N$
that has a
conformal UV fixed point, then the quiver diagram is 
complete. 

The superpotential will be of the form $\sum_{[\alpha]} \lambda_{[\alpha]} 
\tr(\phi^{[\alpha]})$, where $[\alpha]$ is a multi-index in the variables $a$,
 such that
for any two consecutive indices in the operator, the arrow $\phi^{a_i}$ ends 
at the same vertex than where the arrow
$\phi^{a_{i+1}}$ begins, and we contract the gauge indices at the vertex
 between these two 
fields.
This property is true also when we take into account the cyclic 
property of the trace as well.

This is, for each closed loop in the quiver
 diagram (modulo rotations of the initial vertex) there is an associated
superpotential term that we can consider 
adding, but in the end, for most of the interesting theories 
there will only be a finite 
number of $\lambda$ which are different from 
zero.

It is clear that the important aspect of the gauge structure is that
it lets us multiply 
arrows that end at a vertex $V_i$, with arrows that begin at that vertex; 
and the new
composite meson field is also an arrow which transforms as the $(N, \bar N)$
 under two different
gauge groups. Notice that if we want to interpret the arrows
as matrices operating on vectors  by left multiplication,
 the space on which one 
operates is
the second index 
of the arrow, and the space one lands in is associated to
the first index \footnote{Fundamental
 arrows starting at $j$ and ending at 
$i$ represent generators of the group $ext^1([i],[j])$ for the 
coherent sheaves (D-branes) 
associated to nodes $i,j$.
 This is the opposite convention for the arrows 
than \cite{BL4}. See also \cite{Dcat} for how the 
$ext$ groups appear in the discussion}.  
We can take actually any composite meson field 
made with an arbitrary number of arrows and we see
that it also has this property. 
We can thus consider representing 
each fundamental field or meson field  by a
square matrix whose indices are all the possible
gauge indices of the different gauge groups, and where the matrix associated 
to the field
$\phi^a_{ij}$ has entries only in the $[ij]$ block of this big matrix.
The procedure is described in figure \ref{fig:composite.eps}.

\myfig{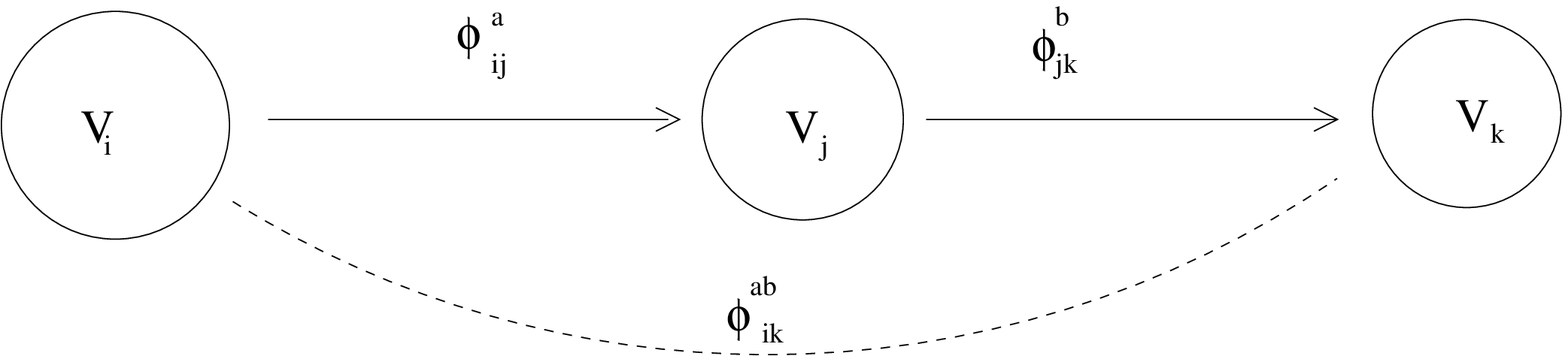}{7}{Composite meson field
in the quiver diagram: $\phi^{ab}_{ik} = \phi^a_{ij}\phi^b_{jl}$}

Any composite meson field as we described can thus be made out 
of multiplying and adding
these types of matrices together, and will also be a matrix in these same 
conventions.
We can call this formal concatenation of symbols a $\star$ operation,
 and it can be seen that it is an associative 
multiplication of matrices once we take into account the contraction
of the gauge indices. The equations of motion derived
from the superpotential are also elements of this algebra. We use these 
equations of motion as relations in the algebra.
 In this way, the algebra encodes the information 
of the moduli space of vacua.

As such, we have a formal algebra of matrices generated by the arrows in 
the quiver, and 
with relations given by the superpotential equations of motion.
 We want the 
gauge group to be
 associated to matrices in this algebra as well, but we need to project it
so that it lies only along the block diagonal elements of the matrix.

To do this, it is convenient to introduce a projector for each of the vertices 
$P_i$, such that $P_i^2 = P_i$, and such that the gauge group is generated 
by matrices of the 
form
\begin{equation}
G = \sum_i P_i G P_i 
\end{equation}
Similarly, the fields $\phi^a$ are such that
\begin{equation}
\phi^a_{ij} = P_i \phi^a P_j = P_i \phi^a_{ij} = \phi^a_{ij} P_j
\label{eq:arrows}
\end{equation}
so these projectors serve to keep track of the blocks in the matrices.
Notice also that $P_i P_j = \delta_{ij} P_j$, so the projectors are mutually
orthogonal. These projectors are not considered as truly dynamical variables,
because they can only have eigenvalues $0,1$ and not arbitrary
complex numbers. 

Notice that $N_i = \tr(P_i)$, 
so we can measure the rank of the gauge groups
 by taking traces of 
these projectors, so they do measure some of the discrete degrees of freedom 
associated to the 
choices we make on a quiver diagram.

The meson algebra $\CA$ associated to the quiver diagram
 will be generated by the 
chiral fields and
the projectors. It is the algebra of meson fields, and it
 is an associative algebra by construction. 
The relations in this algebra
 are given by the fact that
the projectors are mutually orthogonal, by the superpotential equations of 
motion and
by the equations \ref{eq:arrows} which specify where the arrows begin 
and end. The variables in the algebra are precursors of gauge invariant 
operators. To produce a gauge invariant operator we just need to take a trace of
the element in the algebra.

To solve for a point in the moduli space of vacua is then to look for 
solutions to the 
superpotential equations of motion where we have specific matrices for all 
the arrows and projectors.
This amounts to finding a
representation of the non-commutative algebra as given 
above, 
with the projectors included,
up to gauge equivalence. Notice that the main role of the 
projectors is to reduce the gauge
group from the one of $n\times n$ matrices to the smaller group that commutes
 with all of
the $P_i$, and to give the algebra an identity operator
$1 = \sum_i P_i$.

In this way we can take the moduli space problem and phrase it in terms
 of representation 
theory of a generically non-commutative algebra. The algebra is non-commutative
any
 time we have more than one node
as it is easy to see that 
\begin{equation}
P_1\phi_{12} = \phi_{12} \neq \phi_{12} P_1=0 
\end{equation}

The physical interpretation now is that this non-commutative algebra is 
associated to a non-commutative (algebraic) geometry  
that $D$-branes see, 
and we need to extract the closed string target space out of this 
non-commutative geometry. 
Once we have this non-commutative interpretation 
of the quantum field theory, the techniques developed in 
\cite{BJL,BJL2,BL3,BL4}
can be utilized to analyze the field theory moduli space.

An important subalgebra of the non-commutative algebra $\CA$ is the center
 of the algebra, $\ZA$,
which corresponds to  a commutative geometry. 
When the center of the algebra is big enough (we will make this notion 
precise later), this commutative 
geometry gives rise to some singular algebraic geometry 
 which turns out to be of dimension $3$ for most
interesting superconformal models.

The key point of this commutative algebra is the
 idea exposed in the Seiberg-Witten \cite{SW} approach 
to non-commutative field 
theory. There are two geometries: one non-commutative for open strings 
(D-branes), and a commutative 
geometry for closed strings. In the present case, 
the commutative geometry for 
closed strings will correspond to 
the algebraic geometry associated to the
center of the quiver algebra. We will call this variety $V$.

One then needs to show that the moduli space of the D-branes is 
essentially the symmetric space
$Sym^N(V)$. 
The fact that the moduli space is some symmetric space follows from the  
statement that the
direct sum of two 
representations of an associative algebra gives a representation of the 
algebra \cite{BJL}.
Since one can choose the matrices
associated to the direct sum representations to have block diagonal form,
 one can have an 
interpretation of the configuration in terms of two objects
on a given space,  as in matrix theory \cite{BFSS}.
One needs to show  that the space of irreducible representations (which 
are the non-commutative 
analog of points)
away from the 
singularities is
$V$, and that at the singularities 
there are additional fractional brane representations.
The trick that lets us relate the irreducible representations to $V$ is 
Schur's lemma. It tells us that on an irreducible representation of an
 algebra in terms
of $n\times n$ matrices, any element of the center of the algebra is 
proportional to the 
identity. In this way we have an evaluation map from non-commutative points 
to the variety $V$. 

If $\CA$ is finitely generated as a module 
over $\ZA$, then the irreducible 
representations
of $\CA$ have bounded dimension. 
Consider $s_\alpha$, $\alpha= 1,\dots n$
a basis of $\CA$ as a module over 
$\ZA$, where we can take $s_1 = 1$ if we want to.
 Then the relations in the algebra have to read
\begin{equation}
s_\alpha  s_\beta  = \sum_\gamma f_{\alpha\beta\gamma}(\ZA) s_\gamma
\end{equation}
Given a vector $v$ in the representation of the algebra, the
elements $s_\alpha v$ generate the representation, as the 
$f_{\alpha\beta\gamma}$ are given by constant numbers. 

These type of algebras guarantee that a bulk D-brane is made out of
finitely many fractional branes, and it is usually the case
that when this  condition  is  met there are enough representations
in the algebra to cover $V$ completely, as we 
will see in the examples.

If this finitely generated condition is not met, 
then generically one can not cover the
variety $V$ with finite dimensional representations of the quiver,
and therefore the moduli space of vacua does not behave like 
$Sym^n(V)$ for any $n$. This is the statement that the center of the 
algebra is not large enough. Hence we will ask that the quiver
algebra is indeed finitely generated over the center in order to make 
sense of the geometry of $V$ from the moduli space  of vacua.

In the paper \cite{B} it was explicitly verified that these non-commutative
 techniques reproduced the 
conifold geometry exactly for various distinct constructions of the
 non-commutative algebra, including the fractional branes 
at the singularity.

To summarize the construction, here is the recipe for producing the 
reverse geometric engineered complex geometry.

\begin{enumerate}
\item Give a connected 
quiver diagram with some choice of superpotential, such that
there is at least one set of integers $N_i>0$  for which the gauge theory 
associated to the quiver with gauge group $U(N N_i)$ in vertex $V_i$ is 
consistent in
four dimensions (the anomaly factorizes). This is necessary to have a bulk 
D-brane.

\item Construct the quiver algebra $\CA$ with generators 
given by the chiral fields
and with a projector associated to each node. The relations 
in the generators are given by the 
superpotential
equations of motion
and by the location 
of arrows in the quiver diagram.

\item Out of $\CA$ extract the center of the algebra $\ZA$. Verify that
$\ZA$ correspond to the ring of holomorphic functions on a 
variety of complex dimension $3$, which we will call $V$.
For this center to be interesting it is clear that we need to have enough
relations in the quiver algebra to move any arrow from the 
left of an element of the 
center to the right. In particular this implies that every arrow in 
the quiver diagram needs to appear in at
least one term in the superpotential.

\item Verify that for each point in $V$ away from the singularities of $V$ 
there is one  unique irreducible 
representation of the algebra $\CA$. Usually this is 
accomplished only if the algebra $\CA$ is finitely generated 
over $\ZA$. In brane language, this tells us that a bulk brane is 
made out of finitely many fractional branes, so we will require this 
property as a condition to check.

\item If all of the above conditions are met, we say that the field theory
 corresponds to the dynamics of point like (and fractional) 
D-branes on $V$.

\end{enumerate}

{\bf Important Remark:} The construction above provides 
a purely algebraic geometric background. For other models
it might be the case that it is still possible to produce
AdS/CFT duals of the theories (see \cite{BL} for example),
 but these might involve 
turning on  $RR$ and $NS$ two forms. For these models the $RR$
potentials obstruct the moduli space of the D-branes so 
that point like branes can not explore the commutative geometry in it's
entirety. So, although these quiver theories are consistent field theories,
they
lead to a model which 
can not be interpreted in terms of a $(2,2)$ sigma model
on the worldsheet. A second obstruction to the above process of 
defining
a commutative geometry is that we might 
have some set of branes in a quiver diagram which are 
extended when the target space is compact. If these extended branes are such
 that they can not 
all be shrunk to zero geometrical size simultaneously, then the idea that 
we have
all of the fractional branes at a singularity is not valid either.
The rest of the paper will deal with examples where the conditions for finding
the geometry of the variety $V$ are satisfied.

The second point worth noticing, is that the procedure described above is 
not an algorithm. The reason for this is that there is no known algorithm 
to calculate the center of an abstract algebra. 
The ability to calculate the center 
effectively depends on how tractable the structure of the quiver algebra 
is.

\section{Families of resolved $A_{n-1}$ singularities}
\label{sec:fam}
Here we will consider the affine quiver diagrams of an
$A_{n-1}$ singularity with a deformed superpotential, as
has appeared in \cite{CIKV}. See also \cite{OT}. 

The quiver diagram is as shown in the figure \ref{fig:quiver.eps}. We have 
$n$ nodes, $V_{i}$, $i=1,\dots ,n$. Three families of fields
$x_{i,i+1}$, $y_{i,i-1}$ and $z_{i,i}$.

\myfig{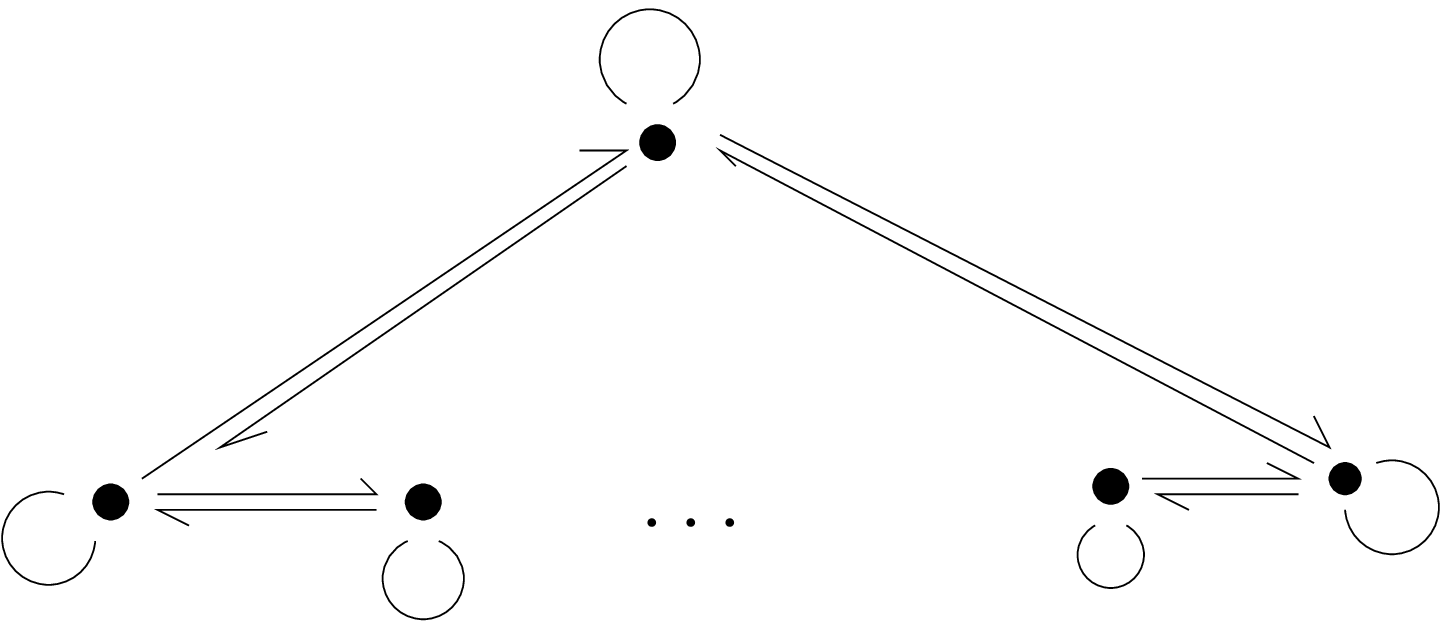}{5}{$A_{n-1}$ quiver diagram}

The superpotential is given by 
\begin{equation}
\sum_i \tr(z_{ii} x _{i,i+1} y_{i+1,i}- z_{ii} y_{i,i-1} x_{i-1,i})
- \tr(Q_i(z_{i,i}))
\end{equation}
and $i = i' \mod(n)$ when the index falls outside ${1,\dots n}$, and 
with the $\beta_i$ polynomials of degree less than or equal to 
$k$ for some $k$,
and such that $\sum_i Q_i(z) =0$, if we replace all the 
variables $z_{i,i}$ by a single variable $z$.

The quiver algebra has projectors $P_i$ for 
$i=1,\dots, n$, and is generated by the $x_{i,i+1}, y_{i,i-1},
z_{ii}$ with additional constraints 
\begin{eqnarray}
x_{i,i+1}y_{i+1,i} - y_{i,i-1} x_{i-1,i} - Q'_i(z_i) &=& 0\\
z_{i,i} x_{i,i+1} - x_{i,i+1} z_{i+1,i+1}&=&0\\
y_{i+1,i} z_{i,i} - z_{i+1,i+1} y_{i+1,i}&=&0 
\end{eqnarray}
and the projection equations $P_j x_{i,i+1} = \delta_{i,j} x_{i,i+1}$
etc.

Consider the new variables
\begin{eqnarray}
z &=& \sum_i z_{ii}\\
x &=& \sum_i x_{i,i+1}\\
y &=& \sum_i y_{i+1,i}\\
\sigma  &=& \sum_i P_i \eta^i
\end{eqnarray}
for $\eta = \exp(2\pi i/n)$ a primitive n-th root of 
unity.

It is easy to see that we can recover the $P_i$ from $\sigma$
when we consider the $n$ monomials $\sigma^k$ for $k=1,\dots,n$.
Namely
\begin{equation}
P_k = \frac1n\sum_j \eta^{-k j} \sigma^j
\end{equation}
From here we can also recuperate the 
individual $x_{i,i+1} = P_i x$, and similarly for $y_{i+1,i}$, 
$z_{i,i}$, so we have a new set of generators of the algebra over 
$\BC$. These are $x,y,z,\sigma$.

In these new variables, the relations read
\begin{eqnarray}
 \sigma z &=& z \sigma\\
 \sigma x &=& \eta x \sigma\\
 \sigma y &=& \eta^{-1} y \sigma\\ 
 x z &=& z x\\
 y z &=& z y\\
 xy -yx &=& \sum_{k=1}^{n-1} \tilde Q_k(z) \sigma^k
\end{eqnarray}
The condition $\sum_k Q_k = 0$ is necessary so that 
on the right hand side there is no term proportional to 
$\sigma^n =1$.  In the expression above 
$\tilde Q$ are some new linear combinations of 
the derivatives of $Q'$. It follows that 
\begin{equation}
Q_j' = P_j  \sum_{k=1}^{n-1} \tilde Q_k(z) \sigma^k P_j
= \sum_{k=1}^{n-1} \tilde Q_k(z_{jj})\eta^{kj}
\end{equation}

It is clear from the equations above that $z$ is in the center of the 
algebra. It is also an  easy matter to check that 
$u= x^n, v= y^n$ are elements
in the center of the algebra. This is necessary in order to commute with 
$\sigma$, and the commutator with $x$ or $y$ is zero from identities obtained
from sums over roots of unity.

There is one additional element of the center. Notice that $xy$ commutes with
 $\sigma$, but it does not commute with $x,y$. Indeed
\begin{equation}
(x) xy - xy (x) = x (xy-yx) = x \sum_k \tilde Q_k(z) \sigma^k
\end{equation}
But we also see that $ x \sigma^ k = \sigma^ k \eta^{-k} x$,
so each of the terms in the right hand side can be written
as a commutator
\begin{equation}
 x \sigma^ k = [x, \sigma^k/(1-\eta^k)]
\end{equation}
and we have
\begin{equation}
[x, xy -  \sum_k \tilde Q_k(z)\sigma^k/(1-\eta^k)] =0 
\end{equation}
We can call this variable $w$,
\begin{equation}
w =  
xy -  \sum_k \tilde Q_k(z)\sigma^k/(1-\eta^k)\label{eq:w}
\end{equation}

Now let us consider the irreducible representations of the algebra.
Clearly, the variable $xy$ is block diagonal, and from \ref{eq:w}
we see that it is proportional to the identity in each block, since $z$ and 
$\sigma $ are too. 
This is, $x_{i,i+1} y_{i+1,i}$ is proportional to the 
identity in the block corresponding to the node $i$, and it is equal
to 
\begin{equation}
x_{i,i+1} y_{i+1,i} = 
P_i w+\sum_k \tilde Q_k P_i \eta^{ik}/(1-\eta^k) 
\end{equation}
So we can consider the quantity $ r_i =  \sum_k \tilde Q_k(z) \eta^{ik}/
(1-\eta^k)$, and we have $ x_{i,i+1} y_{i+1,i} = (w+r_i)P_i$.

If $x^n\neq 0$ and $y^n\neq 0$,  then the product of the $x_{i,i+1}$ 
and $y_{i,i-1}$ is invertible, so each of them is invertible as 
well.
From the fact that $x_{i,i+1} y_{i+1,i}$
is proportional to the identity on the block 
$i$, it follows that all of the nodes have the same rank, this is, 
$x_{i,i+1}$
and $y_{i,i-1}$ establish isomorphisms between the neighboring 
nodes in the diagram. If the rank of each gauge 
group to be bigger than one, let's say $N$,
 then the unbroken gauge group is a diagonal 
$U(N)$, since $x_{i,i+1}$ and $y_{i+1,i}$ are inverses 
of each other, and since $x^n$ and $y^n$ are proportional to the 
identity. Irreducibility implies that the unbroken gauge group is 
$U(1)$, so $N_i=1$ for all $i$. This is the condition for a single brane to
be in the bulk.

Thus in the above, we can treat the $x_{i,i+1}$ and $y_{i+1,i}$
as numbers.

We have then
\begin{eqnarray}
u = x^n &=& \prod_i x_{i,i+1}\\
v = y^ n &=& \prod_i y_{i+1,i}\\
x_i y_i &=& w +r_i
\end{eqnarray}
And we have the relation
\begin{equation}
u v = \prod_i (w+r_i)\label{eq:fibs}
\end{equation}
where each $r_i$ is  polynomial in $z$. This equation is true 
in the full algebra $\CA$. Verifying directly, as opposed to in each of the 
irreducible representations, takes a lot of manipulations with 
algebra, although it is easy to convince oneself from doing the algebra 
explicitly
that it is indeed correct for the
$A_1$ and $A_2$ singularities. Notice also that in the above equation
$\sum r_i =0$.

Here we see that the commutative geometry represents a family of
 resolutions 
of the $A_{n-1}$ singularity, $u v = w^n$, parametrized by $z$. This
is the original geometrically engineered version of the theory in 
\cite{CKV,CIKV}.TThe analysis of these theories
 in terms of brane engineering and M-theory was worked in \cite{
DOT,DOT2,DOT3,OT}

Indeed, it is possible to calculate explicitly the representations. 
We have already seen that $N_i=1$. Notice that in the basis that diagonalizes 
the gauge group, we also have the $\sigma$ diagonal and proportional to 
$\eta^k$
on node $k$. Call these states $|k>$.
Now fix $z$ and $u=x^n\neq 0$ let's say. Since $x^n\neq 0$, it is easy
to check that
$x^m |k> \sim |k+m>$, and we can choose the normalization factor to be 
constant 
by use of gauge transformations; so $x^m |k> = \alpha^m |k+m>$, with 
$\alpha^n = u$.
So the orbit of the variable $x$ acting on $|k>$
is the full representation.
  The equations for the $y_i$ are then linear difference 
equations 
\begin{equation}
y_{i+1,i} - y_{i,i-1} =  \alpha^{-1} Q'_i
\end{equation}
so that all of the $y_{i,i+1}$ are determined once $y_{0,1}$ is known. The condition
$\sum Q'_i(z) = 0$ is then required so that the linearly 
dependent equations above have a solution. 
Given $y_{1.0}$, one determines a unique value for $w$ and $v = y^n$, and vice-versa;
given a value of $w,v$ which satisfies the constraints in the commutative algebra 
defined by $u,v,w,z$, there is a unique solution for $y_{1,0}$ which is compatible 
with it. Notice that in this normalization
$y_{i,i-1}|k> = \tilde y_{i,i-1} |k-1>$, where the
$\tilde y$ is a number.

Now, it is interesting to ask when can the representation be reducible. 
Indeed, from the form of the matrices it is clear that we need to be in a 
situation when applying $x$ $n$ times does not produce a full $n$ dimensional 
representation, and similarly for $y$. This is, we need $x^n = y^n=0$, so there is at
least one $x_{i,i+1}=0$ and one $y_{j+1,j}=0$; and we want these two to be 
associated to $i\neq j$, as otherwise the representation is of dimension 
$n$, and not reducible.

The conditions $x_{i,i+1} = 0$ and $y_{j+1,j} = 0$ mean that
\begin{eqnarray}
x_{i,i+1}y_{i+1,i} &=& P_i(w+r_i) = 0\\
x_{j,j+1}y_{j+1,j} &=& P_j(w+r_j) =0
\end{eqnarray}
So we find that $w +r_i=0$ and $w+r_j=0$,
 since these are variables in the center and $P_{i,j}\neq 0$. This is, the 
representation is reducible exactly when two of the $r_i$ are equal and
$w= -r_i$. These are exactly the points that correspond to the 
singularities in the associated commutative  geometry \ref{eq:fibs},
 when  two of the roots
in the product are equal $r_i = r_j$, $w=-r_i$ is minus the value of 
those equal roots, and when $u=v =0$ too. 

These smaller representations are fractional branes. Indeed, it is easy to see 
that  the fractional brane representations 
correspond to roots of the extended Dynkin digram of the $\hat A_{n-1}$
 singularity,
as they are related to $k$ consecutive nodes having $U(1)$ gauge group, and
$n-k$ consecutive nodes having $U(0)$ gauge group. 

\section{A non-toric non-orbifold singularity}\label{sec:nontor}

Consider the quiver diagram given by one node and three adjoints, similar to 
the ${\cal N = 4}$ quiver diagram, but with a different superpotential
\begin{equation}
\tr( XYZ + ZYX) +\frac\lambda3 (X^3+Y^3+Z^3)
\end{equation}
with $\lambda$ arbitrary. By the arguments of Leigh and Strassler \cite{LS}, the $SU(N)$
gauge theory with this superpotential is conformal (one exploits the $\BZ_3$
symmetry $X\to Y \to Z$) for some value of the gauge coupling.

The quiver algebra has only one projector which is the identity, and the 
constraints are
\begin{eqnarray}
XY +YX &=& \lambda Z^2\\
YZ+ZY &=& \lambda X^2\\
XZ+ZX &=& \lambda Y^2
\end{eqnarray}
Consider the variables $X^2, Y^2, Z^2$. 

It is easy to see that
\begin{equation}
[X^2,Y] = -\lambda [Z^2,X] = \lambda^2 [Y^2,Z] = -\lambda^3 [X^2,Y]
\end{equation}
so that for generic $\lambda $ we have that $X^2$ commutes with $Y$.
One can produce a similar argument that shows that $X^2$ commutes with $Z$. 
And therefore $u= X^2,v= Y^2,w= Z^2$ are all variables in the center of the 
algebra.

It is easy to see that the algebra is finite dimensional over the center, as
we can choose an order where $X$ are before $Y$ and $Y$ are before $Z$ by using the 
equations in the algebra; since the quantities $X^2,Y^2,Z^2$ on the right hand side
of the equations commute with everything.

Consider now
\begin{equation}
[XYZ,X] = XY \lambda Y^2 - X \lambda Z^2 Z
= \frac \lambda2[(XY^3 - Y^3 X) - (XZ^3- Z^3 X)- (X X^3 -X X^3)]
\end{equation}
So both sides are written as a commutator with $X$. 
Thus $ \gamma = XYZ - \frac\lambda2(X^3-Y^3 +Z^3)$ commutes with $X$. 
Notice that this expression is invariant under the $\BZ_3$ group that
takes $X\to Y\to Z$, so it is in the center of the algebra; as it must 
also commute 
with $Y$ and $Z$.

The algebra is of dimension $9$ as a free module
over the center of the algebra, and this 
corresponds to bulk representations in terms of 
$3\times 3$ matrices.

It is now an easy matter to establish, using the same type of manipulations 
as in \cite{B} that
\begin{equation}
\gamma^2 = a uvw + b\lambda^2(u^3+v^3 +w^3)
\end{equation}
for some coefficients $a,b$.
When $\lambda=0$, this variety  corresponds to a $\BC^3/\BZ_2\times \BZ_2$
orbifold, and the field theory describes the singularity with non-trivial 
discrete torsion. 
The variety above for generic $\lambda$ has an isolated singularity
at $\gamma = 0 = u=v=w$. The singularity does not seem to correspond to 
a toric singularity, as the equations satisfied by the variables
are not given in terms of monomials that are equated to each other.
Also, the right hand side of the equation can not be factored into linear
terms
for a generic $\lambda \neq 0$, as this would imply that the curve
in $u,v,w$ is singular inside $\CP^3$, so the singularity would not be
 isolated 
at the origin.

This singularity is obtained by complex structure deformation of a known 
singularity however. Notice that since the generic representation is of dimension 
$3$, this is, we have a $U(3)$ gauge group for the bulk brane, and
 it splits at the origin into three trivial representations of the 
algebra where $Z,Y,X=0$, each with $U(1)$ gauge group.
The fractional branes are identical to one another.

\section{The $\BC^3/\BZ_3$ orbifold and 
Seiberg duality}\label{sec:Z3}

We will now show another example that represents the orbifold $\BC^3/\BZ_3$
after performing a duality on one of the nodes. One of the main reasons to 
understand this problem is from the knowledge that field theories 
which allow fractional branes 
give rise to cascading dualities \cite{KS} and geometric 
transitions. Although this is not the case for the $\BC^3/\BZ_3$ orbifold
because there are no anomaly free configurations with fractional branes,
one can still change the gauge couplings in the theory so that one 
of the nodes
becomes infinitely strongly coupled. The continuation of the theory past this
singular point is a dual theory. However, since this change is affected only 
by the  Khaler structure of the target space, the holomorphic information 
should be invariant
and one should be able to identify the geometric target space of both 
quiver diagrams. 
To set up the problem of the $\BC^3/\BZ_3$ orbifold, let us
 first consider the orbifold quiver of $\BC^3/\BZ_3$ itself.

The orbifold quiver of $\BC^3/\BZ_3$ has three nodes
 $V_{123}$, as depicted in the left diagram of figure 
\ref{fig:z3quiver.eps}, and three sets of arrows 
$\phi^{1,2,3}_{i,i+1}$. The superpotential is given by
\begin{equation}
\tr( \epsilon_{ijk} \phi^i_{1,2} \phi^j_{2,3} \phi^k_{3,1})
\end{equation}
The field theory has an $SU(3)$ global symmetry under which 
each of the $\phi_{i,i+1}$ transforms as a $3$. 

\myfig{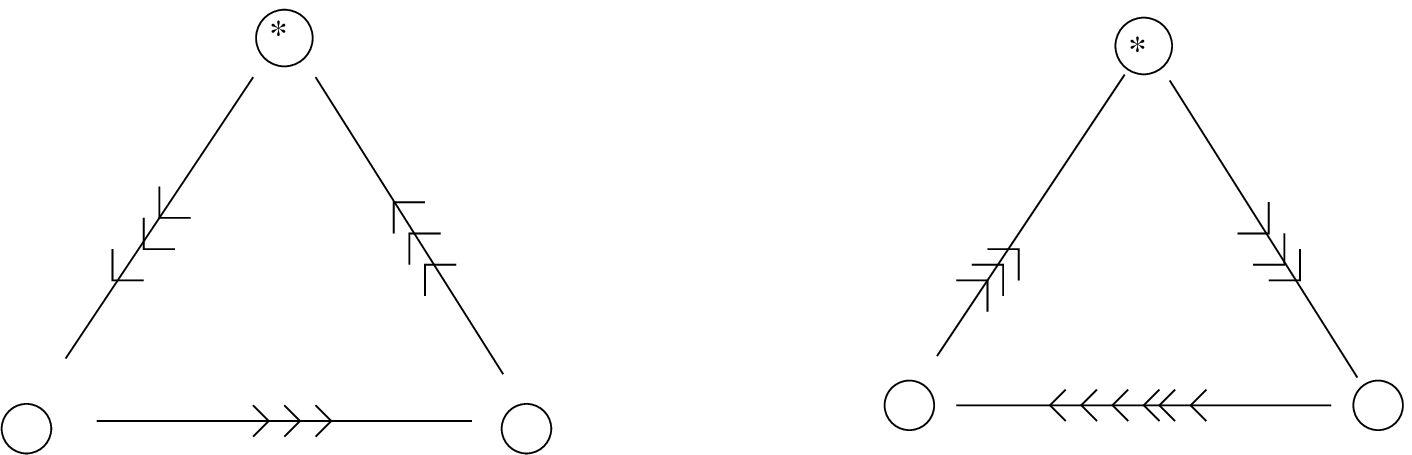}{7}{Two dual versions of the $\BC^3/\BZ_3$ quiver
 diagram, dualized at the marked node}

If we consider the quiver algebra, and the fields $\phi^i$ defined as 
follows
\begin{equation}
\phi^i = \phi^i_{1,2} +\phi^i_{2,3}+\phi^i_{3,1}
\end{equation}
then the superpotential equations of motion reduce to 
\begin{equation}
[\phi^i,\phi^j] =0
\end{equation}
Thus the algebra of the $\phi^i$ is commutative and one can think of
 them as generators of $\BC^3$, however, 
to be elements
of the center of the quiver algebra,
 they need to commute with the projectors associated to the vertices as
well. These elements of the center
are generated by elements of the form 
\begin{equation}
z^{ijk} = \phi^i\phi^j\phi^k
\end{equation}
and one can  see that they transform in the $10$ of the global 
$SU(3)$ (totally symmetric in three $SU(3)$ indices),
 moreover these are exactly the coordinates that are invariant under 
the $\BZ^3$ action on $\BC^3$ that sends $\phi^i \to \eta\phi^i$, where
$\eta^3 = 1$. So the algebra of the center describes the orbifold space 
$\BC^3/\BZ_3$.

Now take the quiver diagram, and perform Seiberg duality \cite{Sei}
on one of 
it's nodes. The resulting quiver diagram is depicted on the right of
 figure
\ref{fig:z3quiver.eps}.
 This does not change the moduli space of vacua,
 so one should be able to
identify $V$ from any of it's dual versions.
The duality for this particular case was described in
\cite{CIKV}, and gives rise to a new quiver diagram where there is no 
apparent orbifold point at which there is an extra $\BZ_3$ symmetry of 
rotating the nodes
clockwise. More general dualities related to toric singularities have been 
analyzed and proposed in \cite{FHH,FHH2,FHHU,BPR}.
 Since it is not clear that after performing Seiberg duality 
the center of the new quiver algebra has any relation with the center of 
the original quiver algebra, it is worth checking that the centers 
give rise to the same commutative geometry. Later we will argue why
this should be the case always.

The quiver diagram to consider is as shown in the figure. We have 
field $\chi^{\alpha\beta}$, $\phi_\alpha$, $\xi_\beta$ which transform
in the $6$, $\bar 3$, $\bar 3$ representations of the  $SU(3)$ global 
symmetry, and we are given the superpotential
\begin{equation}
W = \tr(\chi^{\alpha\beta} \phi_\alpha\xi_\beta)
\end{equation}

We have three projectors $P_{1,2,3}$, and the non-trivial constraints
\begin{eqnarray}
\phi_\alpha\xi_\beta+ \phi_\beta\xi_\alpha &=&0\\
\chi^{\alpha\beta} \phi_\beta &=& 0\\
\xi_\beta \chi^{\alpha\beta} &=&0  
\end{eqnarray}
For a variable to be in the center, it must commute with the three projectors,
and since we have a conformal field theory at the origin in moduli space
let us pick one such of minimal conformal weight which is not the identity. 
It must be of the form 
\begin{equation}
a_1 \chi\xi\phi+ a_2 \phi\chi\xi+ a_3\xi\phi\chi
\end{equation}
This expression we decompose into 
irreducible representations of
the $SU(3)$ global symmetry, and we want to check that there
is a $10$ in the center of the algebra.

Notice that there is only one $10$ dimensional representation of
$SU(3)$ in the product 
\begin{equation}
6\otimes \bar 3\otimes \bar 3
= 1 \oplus 8 \oplus 8 \oplus 27\oplus 10
\end{equation}
which is antisymmetric in the two $\bar 3$ indices. From here it is clear
that the condition of being in the center
will fix the ratios of the coefficients 
$a_1:a_2:a_3$. 

It is clear that the above expression commutes automatically with $P_{1,2,3}$,
so we only need to 
check the commutation relations with $\phi, \xi, \chi$.
Indeed, we only need to do this for only one function in the $10$ 
(the highest weight state), as one can  generate any other 
function in the $10$  by using $SU(3)$ rotations on a single state.
The functions are more explicitly given by 
\begin{equation}
f^{\alpha\beta\epsilon} = \epsilon^{\gamma\delta[ \epsilon}
(a_1\chi^{\alpha\beta]}
\xi_\gamma\phi_\delta+ a_2 \phi_\delta\chi^{\alpha\beta]}\xi_\gamma+
 a_3\xi_\gamma\phi_\delta\chi^{\alpha\beta]})
\end{equation}
where repeated indices are contracted and the square parenthesis indicate that
we take a totally symmetric representation in three indices 
$\epsilon,\alpha,\beta$.
In particular, we will take $\epsilon=\alpha=\beta=1$ to check the 
commutation relations, with arbitrary 
$\phi,\chi,\xi$.

Consider  $[f^{111},\phi_\alpha]$. This is equal to
\begin{equation}
a_2 (\phi_2\chi^{11}\xi_3 -\phi_3\chi^{11}\xi_2)\phi_\alpha
-a_1 \phi_\alpha \chi^{11} (2\xi_3\phi_2)
\end{equation}
Now we have to consider the cases $\alpha=3,2,1$ separately.

For $\alpha = 3$, we have
\begin{eqnarray}
a_2 (\phi_2\chi^{11}\xi_3 -\phi_3\chi^{11}\xi_2)\phi_3
-a_1 \phi_3 \chi^{11} (2\xi_3\phi_2)
&=&\\
a_2(\phi_2\chi^{11}\xi_3\phi_3)-(a_2-2a_1) \phi_3\chi^{11}\xi_2\phi_3&=&
\\ (2a_1-a_2)\phi_3\chi^{11}\xi_2\phi_3&&
\end{eqnarray}
where we have used the superpotential equations repeatedly as
 $\xi_3\phi_3 = 0$, and $\xi_2\phi_3 = -\xi_3\phi_2$.
This expression 
vanishes if $a_2 = 2 a_1$. One can check that the same result is obtained
from commuting with $\phi_2$. 

The case $\alpha=1$ is more involved algebraically.
Here we get

\begin{eqnarray}
 a_2 (\phi_2\chi^{11}\xi_3 -\phi_3\chi^{11}\xi_2)\phi_1
-a_1 \phi_1 \chi^{11} (2\xi_3\phi_2)
&=&\\ -2 a_1 (\phi_2\chi^{11} \xi_1\phi_3 -\phi_3\chi^{11}\xi_1\phi_2)
-a_1\phi_1\chi^{11}(2\xi_3\phi_2)&&
\end{eqnarray}
Now we need to use the superpotential relations
that involve $\chi^{ij} \xi_j = \phi_j\chi^{ij}=0$, so
we obtain
\begin{equation}
-2 a_1 (-\phi_2 \chi^{22}\xi_2\phi_3 +\phi_3\chi^{33}\xi_3\phi_2)
+2a_1(\phi_2\chi^{22}+\phi_3\chi^{33})(\xi_3\phi_2)
\end{equation}
and we see that the terms cancel exactly.

The computations for commuting with $\xi$ are similar, and
one obtains that $a_2 = 2 a_3$ as well. in this way, we have 
fixed the ratio $a_1:a_2:a_3 = 1:2:1$; and the result commutes
with 
both $\phi$ and $\xi$. Last of all, one needs to check that the 
expression commutes with $\chi^{ij}$ as well.

It is easy to check the commutation relations with 
$\chi^{11}$, but for the other $\chi^{ij}$ we need to 
convert the indices that are not equal to $1$ to indices that are equal
 to $1$ by applying the superpotential relations. For example
\begin{eqnarray}
 a_1\chi^{12}\xi_2\phi_3\chi^{11} - a_1\chi^{11}\xi_2\phi_3\chi^{12}&=&\\
 -a_1 \chi^{11}\xi_1\phi_3\chi^{11} +a_1\chi^{11}\xi_3\phi_2\chi^{12}&=&\\
 a_1\chi^{11}\xi_3 \phi_1\chi^{11} - a_1\chi^{11}\xi_3\phi_1\chi^{11}&&
\end{eqnarray}
which vanishes. The verification of the commutation relations for
 the remaining
$\chi^{\alpha\beta}$ are left to the reader as an exercise.

The upshot of the calculation is that there are variables in the $10$ 
dimensional representation of $SU(3)$ which are in the center of the 
algebra. One needs to verify that these also satisfy the appropriate
constraints at the $(f)^2$ level.  That is easier to do by 
calculating the 
moduli space of irreducible representations of the algebra
away from the singularity, but this is exactly the type of information
that is invariant under Seiberg duality, so we will not repeat the 
calculation
of the moduli space here.

Now, one might ask why is the center invariant under Seiberg duality 
transformations? 
The answer lies in representation theory. For a single point like brane, 
variables in the 
center are gauge invariant since they are proportional to the identity. 
This is, they
can be written in terms of their trace multiplied by the identity
operator on the quiver algebra. Since these variables are gauge 
invariant, they should be respected by duality transformations; as well as
 the 
relations that they satisfy, since this is the information that determines
 the moduli 
space of vacua of the supersymmetric theory.

Similarly, given a collection of $N$ 
separated branes, these variables will be block diagonal
on the representation and then
the eigenvalues of the matrices $f$ can be recovered by taking traces 
$\tr(f^k)$. In essence, these variables in the center can be reconstructed
from gauge invariant operators on any brane configuration; and the gauge 
invariant variables are independent of which dual representation of 
the theory one chooses.

\section{Conclusion}

We have seen how given a quiver diagram with some choice of superpotential
we can associate to it
an algebro-geometric space $V$ which is such that the moduli space of
D-branes is essentially the symmetric space $Sym^N(V)$, except for the 
singularities 
of $V$ where fractional branes can appear. We have shown many examples of 
the procedure where we find that $V$ is indeed a
complex variety of dimension $3$.

The construction is based on finding the center of a non-commutative 
meson algebra
associated to the quiver field theory. The relations in the quiver algebra
include the superpotential equations of motion and therefore encode 
the moduli space problem for D-branes.
The space $V$ is the variety associated to the center of this quiver 
algebra. One expects that the reverse geometric engineering 
is sensible only if the algebra of
the quiver is finitely generated as a module over the center,
which seems to be a  severe restriction
on the quiver diagram and the superpotential.

On the cases when it exists, the space $V$ is invariant under 
Seiberg dualities, so the construction of $V$ is unambiguous
for different dual descriptions of the same theory.

It would be very interesting if one could go beyond
 the $U(N)$ gauge groups and
include orientifold constructions as well. However, this is more subtle
as one needs to understand the holomorphic 
properties of an algebra of unoriented strings before one can 
hope to develop these techniques further in this direction.

One can also hope to build new singularities from interesting quiver 
theories and perhaps this problem can result more tractable to provide a 
classification of singularities in complex dimension $3$.

{\bf Acknowledgements:} It is a pleasure to thank R. Corrado,
M .Douglas, S. Katz, I. Klebanov,
R. G. Leigh and C. Vafa for useful correspondence and conversations.
Research supported by DOE grant DE-FG02-90ER40542.

\providecommand{\href}[2]{#2}\begingroup\raggedright\endgroup

\end{document}